\documentstyle[12pt]{article}
\newcommand{\be}{\begin{equation}}
\newcommand{\ee}{\end{equation}}
\newcommand{\bea}{\begin{eqnarray}}
\newcommand{\eea}{\end{eqnarray}}
\newcommand{\brr}{\begin{array}}
\newcommand{\err}{\end{array}}
\newcommand{\bc}{\begin{center}}
\newcommand{\ec}{\end{center}}
\newcommand{\nn}{\nonumber}
\newcommand{\ms}{m(a)}
\newcommand{\msms}{m^{\overline{MS}}(\mu)}
\newcommand{\as}{\alpha_{ s}}
\newcommand{\cf}{\frac{N^2-1}{2N}}
\newcommand{\ep}{\epsilon}
\newcommand{\epp}{\epsilon^{\prime}}
\newcommand{\is}{S^{-1}(p,m(a))}
\newcommand{\isz}{S^{-1}(p,m_0)}
\newcommand{\isc}{S^{-1}(p,m^{\overline{MS}}(\mu))}
\begin{document}
\pagestyle{empty}
\begin{flushright}
ROME prep. 94/1018 \\
CERN-TH.7256/94 \\
\end{flushright}
\centerline{\bf{QUARK MASSES FROM LATTICE QCD
AT THE NEXT-TO-LEADING ORDER}}
\vskip 1cm
\centerline{\bf{C.R. Allton$^{a}$,  M. Ciuchini$^{a,b}$, M. Crisafulli$^{a}$,
E. Franco$^a$,}}
\centerline{ \bf{  V. Lubicz$^a$
 and  G. Martinelli$^{a,c}$ }}
\centerline{}
\centerline{$^a$ Dip. di Fisica,
Universit\`a degli Studi di Roma ``La Sapienza" and}
\centerline{INFN, Sezione di Roma, P.le A. Moro 2, 00185 Rome, Italy. }
\centerline{$^b$ INFN, Sezione Sanit\`a, V.le Regina Elena 299,
00161 Rome, Italy. }
\centerline{$^c$ TH Division, CERN, CH-1211 Geneva 23, Switzerland.}
\centerline{}
\centerline{}
\begin{abstract}
Using the results of  several quenched lattice simulations,
we predict the value of the strange and charm quark masses in the
continuum at
the next-to-leading order, $m^{\overline{MS}}_s(\mu=2\,\, \rm{GeV})=
(128 \pm 18)\,\, \rm{MeV}$ and
$m^{\overline{MS}}_{ch}(\mu=2 \,\,\rm{GeV})=(1.48 \pm 0.28)\,\, \rm{GeV}$.
The errors quoted above have been estimated by taking into account
the original statistical error of the lattice results and the uncertainties
coming from the matching of the lattice to the continuum theory.
A detailed presentation of the relevant formulae at the next-to-leading
order and a discussion of the main sources of errors is also presented.
\end{abstract}
\vskip 3cm
\begin{flushleft}
CERN-TH.7256/94 \\
ROME prep. 94/1018 \\
June 1994
\end{flushleft}

\newpage
\pagestyle{plain}
\setcounter{page}{1}
\section{Introduction}
\label{sec:intro}
Quark masses are parameters of the QCD Lagrangian that cannot be
determined within QCD by theoretical considerations only and cannot be
measured directly since quarks do not appear as physical states.
They are however very important for several reasons. In the framework
of GUTs, quark and lepton masses at low energies are related to
the pattern of symmetry breaking at the grand unification
scale. The best-known example is that, within GUTs, one predicts
$m_b \sim 3 m_\tau$ as a consequence of the $SU(5)$ relation
$m_b =  m_\tau$ at the GUT scale and of strong-interaction
effects \cite{mk1}--\cite{mrc}.
Constraints on quark masses and relations between quark masses
and matrix elements of the Cabibbo--Kobayashi--Maskawa mixing
matrix are also found by using some ansatz on the form of the
quark mass matrix \cite{ans1}--\cite{ansf}.  In this framework,
for example, the matrix element $\vert V_{cb}\vert$ depends, among
other things, on the ratio of the strange to bottom quark mass.
The mass of the strange quark is also a crucial  factor appearing in
the evaluation of matrix elements of penguin and electro-penguin
operators, which enter in the $\Delta I=1/2$ $ K \rightarrow
\pi \pi $ amplitude and in the CP violation parameter $\epp/\ep$.
The values of the light-quark masses are necessary also to estimate the
quark-antiquark condensate $\langle 0 \vert \bar \psi \psi \vert 0 \rangle$
which is the chiral order  parameter of the QCD vacuum and finally the
value of the up quark mass is  fundamental for our understanding of the
strong CP problem.
\par Lattice QCD is in principle able to predict the mass of any quark
by fixing to its experimental value
the mass of a hadron containing a quark with the same flavour.
The quark mass that is directly determined in lattice simulations
is the ``bare" lattice quark mass $m(a)$, which can be converted to the
continuum, renormalized mass $m(\mu)$ through a well-defined
 procedure \cite{gamy}--\cite{boc}.
The conversion factor relating $m(a)$ to $m(\mu)$ can be computed
in perturbation theory. Since the typical scale is of order
$1/a$ (or $\mu$), where $a$ is the lattice spacing, and in current
numerical simulations $a^{-1} \sim 2$--$4$ GeV, we expect small
non-perturbative effects. This would imply an accurate determination of quark
masses at a scale larger than $ a^{-1}$. In this respect lattice QCD is
unique, since other techniques,
like QCD sum rules \cite{becchi}--\cite{domi2}, have to work at much
smaller scales, where higher-order corrections \cite{larin} or non-perturbative
effects \cite{nason} may be rather large.
On the lattice, however, at values of the
lattice strong coupling constant $\beta=6/g^2_L(a)$
currently used ($\beta=6.0-6.4$), perturbative corrections are often
quite sizeable,
due to the presence of ``tadpole" diagrams, which are absent in continuum
perturbation theory, and  may give rise to
large uncertainties. In order to reduce these uncertainties,
 two methods, based on ``boosted" perturbation
theory \cite{lm1,lm2} or on non-perturbative techniques \cite{npr},
have recently been developed and will be used in the following.
\par In this paper, using  values of  quark masses determined in
several numerical simulations \cite{all1}--\cite{ukqcd3}, we predict
the corresponding masses in the continuum and discuss the uncertainties
entailed by the use of perturbation theory at the next-to-leading order.
Following ref. \cite{npr}, we also envisage a procedure to determine
the continuum quark mass, which completely avoids the use of lattice
perturbation theory. Our predictions are essentially limited to the
strange and charm quark masses. Actually, simulations do not have
direct access to very light (up and down) quarks, because of finite-volume
effects. Moreover one expects that the ``quenched" approximation
can introduce uncontrolled systematic errors in this case. Thus we do not
have much to say about the up and down quark masses.
On the other hand, we cannot put the {\it b} quark on the lattice, because it
is necessary to satisfy the condition $m_b a \ll 1$ in order to avoid
huge discretization errors. However,
$m_b$ can be obtained from quarkonia
spectroscopy \cite{qcdsronia,leponia}. An alternative method, based
on the heavy-quark effective theory on the lattice will be presented
elsewhere \cite{chriseio}.
\par In comparison with QCD sum rules, which is at present the only
alternative approach to predicting  quark masses, lattice QCD has
advantages and  disadvantages. The main advantage is that, on the
lattice, one is able to work at scales large enough to avoid higher orders
or non-perturbative effects which plague QCD sum rules calculations.
Moreover, by making use of the proposals of refs.\cite{lm1,lm2}
or \cite{npr},
the error due to the large perturbative corrections encountered on the
lattice can be reduced.
 Finally it is possible to reduce discretization errors,
by using some ``improved" lattice quark action, following the proposal
of Symanzik \cite{hw,sw,impro}.
The main disadvantage of the lattice approach is that essentially all the
results have been obtained in the ``quenched" approximation and it is
difficult to
quantify the size of the error introduced by this approximation.
\par
We believe that the results of this work are
useful as a presentation of the method, for the discussion of the
uncertainties coming from the matching of the lattice to the continuum
theory and for phenomenological applications.
 The plan of the paper is the following.
In sec. \ref{sec:latcon} we give the formulae necessary to relate
matrix elements of lattice operators and masses to the corresponding
 quantities in the continuum. The formulae introduced
in sec. \ref{sec:latcon} are applied to quark masses
in sec. \ref{sec:ms}, where lattice perturbation theory and
non-perturbative methods based on Ward identities are reviewed.
Numerical results and estimates of errors are given
in sec. \ref{sec:bk}.
\section{Bare and renormalized operators  on the lattice}
\label{sec:latcon}
In this section we summarize the main formulae that relate  lattice
operators to the corresponding continuum renormalized
operators\footnote{ A more
detailed discussion on this point
can be found in ref. \cite{upgraded}.}. These formulae
will be used to give the continuum renormalized quark mass in terms
of the lattice bare mass, as computed in Monte Carlo simulations.
\par
Let us consider the renormalization of a generic
lattice two-quark operator
of the form $O_\Gamma(a)=
\bar \psi \Gamma \psi$, where  $\Gamma$ is one of the Dirac
matrices and $\psi$ is the bare lattice fermion field.
At the next-to-leading order (NLO)
the generic, forward, two-point Green function,
  computed between quark states of virtuality $p^2 \ll 1/a^2$, has the
form\footnote{ We work on
a Euclidean lattice.}:
\bea \Gamma(pa)& =&
 \Gamma_0 \Bigl[ 1 + \frac{g^2_L(a)}{16 \pi^2}
(\frac{ \gamma^{(0)}}{2} \ln( \frac{p a }{\pi})^2 +C^L ) \nn \\
&+&\Bigl(\frac{g^2_L(a)}{16 \pi^2}\Bigr)^2
\Bigl( \frac{1}{8} \gamma^{(0)}
( -2 \beta_0 + \gamma^{(0)})
\ln^2 (\frac{p a }{\pi})^2 \nn \\  &+&
\frac{1}{2} \Bigl( \bar \gamma^{(1)}+
 ( -2 \beta_0 + \gamma^{(0)}) C^L
+\beta^0_\lambda (\lambda \frac{\partial C^L}{\partial \lambda}
) \Bigr) \ln (\frac{p a }{\pi})^2
\Bigr) \Bigr]  + \dots \label{eq:rgel} \eea
where $\dots$ represent terms beyond the next-to-leading order and
terms of $O(a)$, which we will assume negligible in the following;
$\Gamma_0$ is the zeroth-order Green function, and  $g^2_L(a)=6/\beta$ is the
bare lattice coupling constant; $\gamma^{(0)}$ and
$\bar \gamma^{(1)}$ are the leading (regularization-independent) and
next-to-leading (regularization-dependent)
order anomalous dimensions, respectively;
$\lambda=\lambda(a)$ is the lattice gauge parameter
of the gluon propagator $\Bigl( \Pi_{\mu \nu}(q^2)= -\delta_{\mu \nu}
+(1-\lambda) q_\mu q_\nu/q^2 \Bigr)$. It obeys the renormalization
group equation:
\be \frac{1}{\lambda(a)}  a \frac{\partial \lambda(a)}
{\partial a}=-\beta_\lambda(g^2_L(a))=
-\beta^0_\lambda \frac{g^2_L(a)}{16\pi^2}+\cdots, \,\,\,\,
\,\,\,\,\, \beta^0_\lambda=\frac{5 N-2 n_f}{6\pi}. \label{eq:bl} \ee
We have introduced a scale-dependent gauge parameter in order to define a
gauge-independent anomalous dimension and simplify the renormalization
group equation for $\Gamma(p a)$, see eq. (\ref{eq:rga})  below.
\par The lattice coupling constant obeys the equation:
\be a \frac { dg^2_L(a) } {d a} =- \beta (g^2_L(a) )=
2 \beta_0 \frac{g^4_L(a)}{16\pi^2}
 +2 \beta_1 \frac{g^6_L(a)}{(16\pi^2)^2}, \label{eq:rcc}\ee
where $\beta_{0,1}$ are given by:
\be \beta_0 =\frac {(11N-2 n_f)} {3},  \,\,\,\,\,\,\,\,\,\,\,
\beta_1 = \frac {34}{3} N^2 - \frac{10}{3} N n_f -\frac
{(N^2-1)}{N} n_f \ee
and $n_f$ is the number of flavours.
Equation (\ref{eq:rgel}) guarantees that all the  matrix elements
can be made finite
(as $a \rightarrow 0$) by  multiplying  the bare operator
by a suitable renormalization
constant, obtained by   fixing  the renormalization conditions
for $O_\Gamma$.

\par From the above equations we find:
\be \Bigl( a \frac{\partial}{\partial a} - \beta(g^2_L(a))\frac
{\partial}{\partial g^2_L(a)}-\beta_\lambda(g^2_L(a))
\lambda\frac{\partial}{\partial \lambda}
 -\bar \gamma(g^2_L(a)) \Bigr) \Gamma(p a) =0
\label{eq:rga} \ee
with
\be \bar \gamma(g^2_L(a)) =  \frac{g^2_L(a)}{16\pi^2} \gamma^{(0)}
 + \frac{g^4_L(a)}{(16\pi^2)^2}\bar \gamma^{(1)}. \label{gcc}\ee
\par In view of the comparison with some continuum regularization,
it is convenient to expand the bare Green function $\Gamma(p a)$
of eq. (\ref{eq:rgel}) in terms of the continuum minimal subtraction
($\overline{MS}$)  coupling constant, evaluated at the scale $\pi/a$.
The continuum $\overline{MS}$ coupling  $\as(\pi/a)$ is related to
 the lattice bare  coupling
$\as^L(a)=g^2_L(a)/4 \pi$ by the equation:
\be \frac{1}{\as^L(a)}=\frac{1}{\as(\pi/a)}
\Bigl( 1 + \frac{\as(\pi/a)}{4\pi} \Delta+\dots \Bigr), \label{eq:ras} \ee
where $\Delta$ is a numerical constant.
With this substitution, eq. (\ref{eq:rga}) becomes:
\be \Bigl( a \frac{\partial}{\partial a} - \beta(\as)\frac
{\partial}{\partial \as}-\beta_\lambda(\as)
\lambda\frac{\partial}{\partial \lambda}
 -\gamma_L(\as) \Bigr) \Gamma(p a) =0
\label{eq:rga1} \ee
and
\be \beta(\as) = - 2 \beta_0 \frac{\as^2}{4\pi}
 -2 \beta_1  \frac{\as^3}{(4\pi)^2}, \,\,\,\,\,\,\,\,\,\,
 \gamma_L(\as) =  \frac{\as}{4\pi}\gamma^{(0)}
 +  \frac{\as^2}{(4\pi)^2}\gamma^{(1)}_L, \label{eq:gccb}\ee
\be \beta_\lambda(\as)=\beta^0_\lambda \frac{\as}{4 \pi}+\cdots, \nn
\ee
By changing the expansion parameter, one also has to change the
two loop anomalous dimension \cite{alta}--\cite{ciuc}, see also
eq. (\ref{eq:rgel}):
\be \gamma^{(1)}_L=\bar \gamma^{(1)} -  \Delta \gamma^{(0)} \label{eq:rg1}
\ee
Finally the running coupling constant $\as$ is given by:
\bea \frac {\as(\mu^2)}{4 \pi} =
\frac {1} {\beta_0 ln(\mu^2/\Lambda_{QCD}^2)} \Bigl(
1 - \frac{\beta_1 ln[ln(\mu^2/\Lambda_{QCD}^2)]}{\beta_0^2
ln(\mu^2/\Lambda_{QCD}^2)}\Bigr) + \cdots \label{eq:srcc} \eea
The above equation defines
the continuum $\overline{MS}$ scale parameter $\Lambda_{QCD}$ at the NLO.
\par In continuum dimensional regularizations, after  subtraction of
the poles in $1/\epsilon$,
the renormalized Green function has a form similar to eq. (\ref{eq:rgel}):
\bea&\,& \Gamma(\frac{p}{\mu} ) =
 \Gamma_0 \Bigl[ 1 + \frac{\as(\mu)}{4 \pi}
(\frac{1}{2} \gamma^{(0)} \ln( \frac{p}{\mu  })^2 +C^C ) \nn \\
&+&\Bigl(\frac{\as(\mu)}{4 \pi}\Bigr)^2
\Bigl( \frac{1}{8} \gamma^{(0)}
( -2 \beta_0 + \gamma^{(0)})
\ln^2 (\frac{p}{\mu  })^2 +
\frac{1}{2} \Bigl(  \gamma^{(1)}\nn \\ &+&
 ( -2 \beta_0 + \gamma^{(0)}) C^C
+\beta^0_\lambda (\lambda \frac{\partial C^C}{\partial \lambda}
) \Bigr) \ln( \frac{p}{\mu  })^2 \Bigr]  + \dots \label{eq:rgec} \eea
It obeys the renormalization group equation:
\be \Bigl( \mu \frac{\partial}{\partial \mu} + \beta(\as)\frac
{\partial}{\partial \as} + \beta_\lambda(\as)\lambda
\frac{\partial}{\partial \lambda} +\gamma(\as) \Bigr) \Gamma(\frac{p}{\mu}) =0
, \label{eq:rga2} \ee
with
\be \gamma(\as) =  \frac{\as}{4\pi}\gamma^{(0)}
 +  \frac{\as^2}{(4\pi)^2}\gamma^{(1)}\label{gccc}\nn \ee
and
\be\mu  \frac{\partial\as(\mu)}{\partial \mu}  = \beta(\as(\mu)) ,
\,\,\,\,\,\,\,\,\,\,\, \frac{1}{\lambda(\mu)} \mu
 \frac{\partial\lambda(\mu)}{\partial \mu}= \beta_\lambda(\as). \ee
By solving the renormalization group equations
 for $\Gamma(pa)$ and $\Gamma(p/\mu)$, and imposing
the matching condition $\Gamma(p/\mu)=
C(\mu a) \Gamma(pa)$, we find \cite{upgraded}:
\bea O_\Gamma(\mu)&=& C(\mu a) O_\Gamma(a) \nn \\ &=&
\Bigl(\frac{\as(\mu)}{\as(\pi/a)} \Bigr)^{\gamma^{(0)}/2 \beta_0}
\Bigl[ 1 +\frac{\as(\mu)-\as(\pi/a)}{4\pi}
 \frac{\gamma^{(1)}\beta_0-
\gamma^{(0)}\beta_1}{2 \beta_0^2}  \nn \\ &+&
\frac{\as(\pi/a)}{4\pi} \Bigl(
 C^C-C^L \Bigr) \Bigr] O_\Gamma(a) \nn \\ &\simeq&
\Bigl(\frac{\as(\mu)}{\as(\pi/a)} \Bigr)^{\gamma^{(0)}/2 \beta_0}
\Bigl(1 +\frac{\as(\mu)-\as(\pi/a)}{4\pi}
 \frac{\gamma^{(1)}\beta_0-
\gamma^{(0)}\beta_1}{2 \beta_0^2}\Bigr)  \nn \\ &\times&
\Bigl[ 1 +\frac{\as(\pi/a)}{4\pi} \Bigl(
 C^C-C^L \Bigr) \Bigr] O_\Gamma(a)
 \label{eq:cmw1} \eea
Equation (\ref{eq:cmw1}) has been obtained using the universality of the
combination\footnote{ This relation is
true only when we use the same coupling constant and renormalized
gauge parameter on the lattice and in
 the continuum.}:
\be \frac{\gamma^{(1)}}{2 \beta_0}-C^C=
\frac{\gamma^{(1)}_L}{2 \beta_0}-C^L \label{eq:uni}. \ee
Equation  (\ref{eq:cmw1}) is interpreted as follows: the lattice operator
is matched into the continuum operator at the scale $\pi/a$ through the
factor $1+\as(\pi/a)/4\pi$  $ \Bigl( C^C-C^L \Bigr) $  and then evolved
in the continuum,
according to eq. (\ref{eq:rga2}), from the scale $\pi/a$ up to
$\mu$.
A few comments may be useful at this point:
\begin{itemize}
\item One can eliminate the coefficient $C^L$ in eq. (\ref{eq:cmw1})
by defining a new lattice operator
\be O^{\prime}_\Gamma(a)=O_\Gamma(a)
(1-\frac{\as(\pi/a)}{4\pi} C^L). \ee
This is equivalent  to the regularization independent definition
of the renormalized operators discussed in refs.\cite{alta}--\cite{ciuc}.
This definition is such that
$\Gamma(p=\pi/a)/\Gamma_0=1+O(\as^2)$\footnote{ This is not in contrast with
the condition $p^2 \ll (\pi/a)^2$ that we have to impose in order
to avoid discretization errors. The renormalization condition is simply
a consequence of eq. (\ref{eq:rgel}), which is derived by expanding
in $\as(\pi/a)$ for $p^2 \ll (\pi/a)^2$.}. Notice however
that this procedure requires a  knowledge of the external states for
which we have computed $C^L$ and that it is in general gauge-dependent.
\item We have decided to expand the lattice Green function in $\as(\pi/a)$.
However we have the freedom to expand in $\as(1/a)$ or any other scale
we like since the change will be  completely  compensated at the NLO.
For the same reason we can expand the Green functions
in a different coupling constant,
for example the ``boosted"  coupling $\as^V$ defined
in refs.\cite{lm1,lm2}.  There, it
was argued that the series in $\as^V$ may  minimize $O(\as^2)$ NNLO
corrections. If we expand in $\as^V$ we have accordingly
to reorganize the expression  used for the coefficient function.
It is wise, however, to check the stability of the physical amplitudes under
a change in the scale used for $\as$ and assume this as a theoretical
uncertainty.
\end{itemize}
\section{Quark masses}
\label{sec:ms}
\subsection{ Standard perturbative approach}
\label{sec:ms1}
\par From lattice simulations, by fixing the value of the
lattice spacing and
the physical mass of a strange or charmed hadron,
it is possible to evaluate the bare lattice quark mass $m(a)=m_s(a)$ or
$m_{ch}(a)$ (strange and charmed quark masses).  From $\ms$
one can compute  the quark mass $\msms$, renormalized in
the continuum minimal-subtraction dimensional scheme, at the NLO;
$\msms$ will be defined below.
In the following we report the relevant formulae used
 to relate $\msms$ to $\ms$
and discuss the different sources of theoretical uncertainty. The final
values and errors of  $\msms$ for different flavours
  have been obtained by using the
results of refs.\cite{all1}--\cite{ukqcd3}, as shown in table \ref{tab:ms},
and include the theoretical uncertainties discussed below.
\pagestyle{empty}
\newpage
\pagestyle{empty}

\begin{table}
\begin{center}
\begin{tabular}{||c|c|c|c|c|c|c||}
%
\hline
\rm{Action}
&$\beta$&$a^{-1}$   & $K_c$& $K_s$  & $m_s(a)$
& $\bar m_s(a)$  \\ \hline \hline
Wilson \cite{all1} & 6.0  & $2.23(5)^a$ & 0.15704(3)  & 0.1548(1)&
$0.104(2)$ & $0.127(3)$  \\ \hline
Wilson \cite{all1} & 6.0  & $2.21(8)^b$ & 0.15704(3)  & 0.1547(2)&
$0.105(4)$ & $0.128(4)$  \\ \hline
Wilson \cite{bernard} & 6.0  & $2.29(37)^b$ & 0.1570(1)  & 0.1553(8)&
$0.080(40)$ & $0.098(49)$  \\ \hline
Clover \cite{all1}  & 6.0  & $2.05(6)^a$ & 0.14551(3) & 0.1437(1) &
$0.090(3)$ & $0.103(3)$       \\ \hline
Clover \cite{all1}  & 6.0  & $2.11(11)^b$ & 0.14551(4) &  0.1438(2)
& $0.088(4)$ & $0.100(5)$ \\ \hline
Clover \cite{all3}  & 6.0  & $1.95(7)^a$ & 0.14548(2) &  0.1434(1)
& $0.096(3)$ & $0.108(4)$ \\ \hline
Clover \cite{all3}  & 6.0  & $1.78(9)^b$ & 0.14548(2) &  0.1430(2)
& $0.104(5)$ & $0.117(6)$ \\ \hline
Clover \cite{ukqcd0} & 6.0  & $2.0^{+3~a}_{-2}$ & 0.14556(6) &
$0.1437^{+4}_{-5}$ & $0.089(28)$ & $0.101(31)$  \\ \hline
Wilson \cite{dallas1} & 6.2  & $2.9(2)^a$ & $0.15330(2)$ & $0.1517(3)$ &
$0.101(9) $ & $0.121(10)$  \\ \hline
Wilson \cite{dallas1} & 6.2  & $3.0(2)^b$ & $0.15330(2)$ & $0.1518(2)$ &
$0.098(8) $ & $0.118(9)$  \\ \hline
Wilson \cite{ukqcd2}   & 6.2 & $2.77 ^{+9~a}_{-23}$ & $0.15328 ^{+7}_{-4}$
& $0.1517 ^{+1}_{-3}$ & $0.094(20)$ & $0.113(24)$ \\ \hline
Clover \cite{ukqcd1}  & 6.2  & $2.7(1)^a$ &  0.14315(2) & 0.1419(1)
& $0.082(8)$ & $0.093(8)$ \\ \hline
Clover \cite{all2} & 6.2  & $3.05(19)^a$   & 0.14313(2)  & 0.1421(1)
& $0.079(5)$ & $0.090(5)$ \\ \hline
Clover \cite{all2} & 6.2  & $2.73(17)^b$   & 0.14313(2)  & 0.1418(2)
& $0.089(6)$ & $0.100(7)$ \\ \hline
Wilson \cite{bernard} & 6.3  & $3.01(26)^b$ & 0.15163(3) & 0.1503(3)
& $0.088(21)$ & $0.105(25)$  \\ \hline
Wilson \cite{abada2} & 6.4 & $3.7(1)^a$ & 0.1506(1) & 0.1495(1) &
$0.089(3)$ & $0.106(3)$  \\ \hline
Wilson \cite{abada2} & 6.4 & $4.0(6)^b$ & 0.1506(1) & 0.1497(2) &
$0.082(13)$ & $0.098(15)$  \\ \hline
%
%
%
\hline
\rm{Action}
&$\beta$&$a^{-1}$   & $K_c$& $K_{ch}$  & $m_{ch}(a)$
& $\bar m_{ch}(a)$  \\
\hline \hline
Wilson \cite{dallas1} & 6.0  & $2.18(9)^a$ & 0.15702(4) & 0.1313(16) &
$1.36(4) $ & $1.26(2)$  \\ \hline
Wilson \cite{dallas1} & 6.0  & $1.97(8)^b$ & 0.15702(4) & 0.1271(17) &
$1.48(5) $ & $1.31(2)$  \\ \hline
Clover \cite{dallas1} & 6.0  & $1.92(11)^a$ & $0.14544(2)$ & 0.1212(24) &
$1.32(8) $ & $1.13(3)$  \\ \hline
Clover \cite{dallas1} & 6.0  & $1.94(5)^b$ & $0.14544(2)$ & 0.1218(11) &
$1.30(4) $ & $1.12(2)$  \\ \hline
Wilson \cite{dallas1} & 6.2  & $2.9(2)^a$ & $0.15330(2)$ & $0.1354(24)$ &
$1.24(8) $ & $1.22(5)$  \\ \hline
Wilson \cite{dallas1} & 6.2  & $3.0(2)^b$ & $0.15330(2)$ & $0.1362(22)$ &
$1.22(8) $ & $1.21(5)$  \\ \hline
Wilson \cite{ukqcd3}   & 6.2  & $2.77^{+9~a}_{-23}$ & $0.15328 ^{+7}_{-4}$
& $0.1350(-)$ & $1.22(10)$ & $1.20(10)$ \\ \hline
Clover \cite{ukqcd3}  & 6.2  & $2.6(2)^a$ & $0.14313^{+7}_{-4}$ & $0.1290(-)$
& $0.99(8)$ & $0.94(7)$ \\ \hline
Wilson \cite{abada2} & 6.4 & $3.7(1)^a$ & 0.1506(1) & 0.1386(7) &
$1.06(2)$ & $1.10(2)$  \\ \hline
Wilson \cite{abada2} & 6.4 & $4.0(6)^b$ & 0.1506(1) & 0.1402(24) &
$1.00(11)$ & $1.05(9)$  \\ \hline
\end{tabular}
\caption[]{ \it{Lattice results for the strange and charm quark masses
from lattice QCD. In the table $\beta=6.0/g_L^2(a)$, $K_q$ ($q=s,ch$)
is the lattice hopping parameter for the quark $q$,
$am_q(a)=1/2\Bigl( 1/K_q - 1/K_c \Bigr)$ and $ a\bar m_q(a)=\ln(4 K_c/K_q-3)$.
$a^{-1}$, $m(a)$ and $\bar m(a)$ are given
in GeV.
The symbol $^a$ denotes the scale taken from $M_{\rho}$ and $^b$ from
$f_\pi$. In the second case, we used
 boosted perturbation theory for the renormalization constant
$Z_A$ in the Wilson case ($Z_A=0.78,0.79,0.81$ at $\beta=6.0,6.2,6.4$),
 and the non-perturbative values in the SW--Clover
case ($Z_A=1.09,1.04$
at $\beta=6.0,6.2$ \cite{vla,za}).
Ref. \cite{bernard} used $Z_A=0.77,0.79$ at $\beta=6.0,6.3$,
respectively.}}
\label{tab:ms}
\end{center}
\end{table}
\newpage
\pagestyle{plain}
For definiteness, the formulae used in this section are valid for
the lattice Wilson \cite{wilson} and SW--Clover \cite{sw} quark actions.
The case of staggered fermions will not be considered here.
\par In perturbation theory,
the inverse quark propagator on the lattice can be written as:
\bea \isz = i p\!\!\!/ \Bigl[1-\Sigma_1^L\Bigl(pa,m_0 a)\Bigr]
 + m_0\Bigl[1- \Sigma_2^L(pa,m_0 a)\Bigr]+
\frac{\Sigma_0^L}{a}, \label{eq:qp} \eea
where $a$ is the lattice spacing.
The last term in eq. (\ref{eq:qp}) diverges linearly
as $a \rightarrow 0$. This term  is induced by  the explicit chiral symmetry
breaking that is present in the lattice formulation of the quark action.
The linear divergence is a mass term, which  can be eliminated
by a suitable redefinition of the bare quark mass \cite{gamy}-\cite{boc},
\cite{ks,ka}.
At  $O(\as)$, this is obtained by writing $\isz$ as follows:
\bea \is=  \Bigl[1-\Sigma_1^L(pa,m(a) a)\Bigr]
\nn \\ \times \Bigl[
i p\!\!\!/ + m(a)\Bigl(1+\Sigma_1^L(pa, m(a)a)
- \Sigma_2^L(pa,m(a)a)\Bigr)\Bigr], \label{eq:qp1} \eea
where $m(a)=m_0+\Sigma^L_0/a$.
By comparing eq. (\ref{eq:qp1}) to the corresponding expression in the
continuum:
\bea \isc=  \Bigl[1-\Sigma_1^C(p/\mu, \msms/\mu)\Bigr]
\nn \\ \times \Bigl[
i p\!\!\!/ + \msms\Bigl(1+\Sigma_1^C(p/\mu,\msms/\mu)
 - \Sigma_2^C(p/\mu,\msms/\mu)\Bigr)\Bigr], \label{eq:qpc} \eea
we obtain:
\bea m^{\overline{MS}}(\mu) =m(a) \Bigl[1+\Sigma_1^L(pa,m(a)a)
- \Sigma_2^L(pa,m(a)a)\nn \\- \Sigma_1^C(p/\mu,\msms/\mu)
+ \Sigma_2^C(p/\mu,\msms/\mu) \Bigr] \nn \\
\simeq m(a) \Bigl[1 +\frac{\as}{4 \pi}  \Bigl( \gamma^{(0)}
 \ln(\frac{\pi}{\mu a}) + K_m \Bigr) \Bigr]; \label{eq:mmuma} \eea
$\gamma^{(0)}$ and $K_m$ are  numerical constants, which can be computed from
the general expressions of $\Sigma_{1,2}$:
\bea \Sigma_1^{L,C}(p/\mu,m/\mu)&=& \frac{\as}{4\pi} \cf \Bigl[ \sigma_1^{L,C}
+\int^1_0 dx \Bigl( 2 (1-x) \ln( (p^2 x (1-x)+m^2 x)/\mu^2) \Bigr)\Bigr] \nn \\
\Sigma_2^{L,C}(p/\mu,m/\mu)&=& \frac{\as}{4\pi} \cf \Bigl[ \sigma_2^{L,C}+
\int^1_0 dx
\Bigl( 4 \ln(( p^2 x (1-x)+m^2x)/\mu^2 ) \Bigr)\Bigr] \label{eq:sigs} \eea
where the constants $\sigma^{L,C}_{1,2}$ are reported in table
\ref{tab:sigs}, together with $\Sigma_0^L$, which was defined in
eq. (\ref{eq:qp}).
\begin{table}
\begin{center}
\begin{tabular}{||c|c|c|c|c||} \hline \hline
\rm{Regularization}
  & $\sigma_1$&$\bar \sigma_1$   & $\sigma_2$   & $\Sigma_0^L$     \\ \hline
\hline & & & & \\
$\overline{MS}$      &  1  &$-$& 2  & $-$               \\ \hline
Lattice-Wilson       & 13.85  & 0.99& 1.90  & 51.43         \\ \hline
Lattice-Clover       & 10.11  &2.12     &$ -8.07$  & 31.98       \\ \hline
\hline
\end{tabular}
\caption[]{ \it{Quantities  entering in the definition of $K_m$,
eq. (\ref{eq:relmas});
$\Sigma_0^L$, defined in eq. (\ref{eq:qp}), is given in units of
$(\as/4 \pi) (N^2-1)/2N  $. They have been computed in
refs.\cite{gamy}--\cite{boc},\cite{smit},\cite{frezzo}.}}
\label{tab:sigs}
\end{center}
\end{table}
One finds:
\be \gamma^{(0)}=6  \cf\label{eq:g0}  \ee
and
\be K_m=
 \cf \Bigl( \sigma_1^L-\sigma_2^L-\sigma_1^C+\sigma_2^C -6
\ln(\pi)\Bigr)
= C_m^C-C_m^L .
\label{eq:relmas} \ee
{}From eq. (\ref{eq:mmuma}) we can derive the relation between $\msms$
and $m(a)$ at the next-to-leading order, cf. eq. (\ref{eq:cmw1})
in sec. \ref{sec:latcon}:
\bea \msms= \Bigl(\frac{\as(\mu)}{\as(\pi/a)}\Bigr)^{\gamma^{(0)}/2 \beta_0}
\Bigl[ 1+ \frac{\as(\mu)-\as(\pi/a)}{4 \pi}
\Bigl(  \frac{\gamma^{(1)}}{2\beta_0}-
\frac{\gamma^{(0)}\beta_1}{2\beta_0^2})
\nn \\ +\frac{\as(\pi/a)}{4 \pi} K_m \Bigr] m(a); \label {eq:msren} \eea
$m(a)$, $\msms$, $\gamma^{(1)}$ and $K_m$ are regularization-dependent,
gauge-invariant quantities;
 $\gamma^{(0)}$ is the LO anomalous dimension and $\gamma^{(1)}$
the NLO one, computed in the $\overline{MS}$ dimensional scheme;
$\gamma^{(1)}$ is connected to the lattice anomalous dimension
$\gamma^{(1)}_L$ through the relation, see also eq. (\ref{eq:uni}):
\be  \gamma^{(1)}=\gamma^{(1)}_L+
2 \beta_0 K_m . \ee
 In ref. \cite{douglas}-\cite{nac} they found:
\be \gamma^{(1)}=\frac{97}{3}N \cf
+3 (\cf)^2 -\frac{10 n_f}{3}\cf \label{eq:g1} \ee
In eq. (\ref{eq:msren}),
 $\as$ is the running coupling constant in the continuum
minimal-subtraction scheme.
 It is related to the lattice bare  coupling
$\as^L(a)=g^2_L(a)/4 \pi$ by the equation, see (\ref{eq:ras}):
\be \frac{1}{\as^L(a)}=\frac{1}{\as(q)}
\Bigl( 1+ \frac{\as(q)}{4\pi}
(\beta_0 \ln(\frac{\pi}{qa})^2 + 48.76 )\Bigr)
.\label{eq:asms} \ee
At $\beta=6.0$ one finds:
\be \as(q=\frac{\pi}{a}) \simeq 1.45\,\, \as^L(1/a). \ee
The above definition gives values of $\as$
close to the   values
of the ``optimized perturbative lattice expansion" couplings
$\as^V$ introduced in refs.\cite{lm1,lm2}.
One typically finds $\as^V(\pi/a) \simeq 1.6$--$1.8\,\, \as^L(a)$.
Among the definitions used in refs.\cite{lm1,lm2},
we have:
\be \frac{1}{\as^L(a)}=\frac{1}{\as^{V_1}(q)}
\Bigl( 1+\frac{ \as^{V_1}(q)}{4\pi}
(\beta_0 \ln(\frac{\pi}{qa})^2 + 59.09 )\Bigr) ,
\label{eq:plaq1}\ee
where $\as^{V_1}$ is defined from the $Q$--$\bar Q$
 heavy-quark static potential.
Using the perturbative expansion of the plaquette, one finds:
\be \frac{1}{\as^L(a)}=\frac{1}{\as^{V_1}(q=\pi/a)
\langle \frac{1}{3}{\rm Tr} U_{plaq}\rangle} \Bigl( 1 +\frac{
\as^{V_1}(q=\pi/a)}
{4 \pi} 6.45\Bigr) .
\label{eq:plaq2} \ee
One can expand in $\as^{V_1}$ by
redefining $\gamma^{(1)}$ according to  eqs.(\ref{eq:ras})
and (\ref{eq:rg1}), and use the value of $\as^{V_1}$,
found by using the plaquette computed in numerical simulations.
Another commonly used definition of $\as^V$ is:
\be \frac{1}{\as^L(a)}=\frac{(8 K_c)^4}{\as^{V_2}}\simeq
\frac{1}{\as^{V_2}}(1+  \Sigma_0^L), \label{eq:kckc} \ee
We will discuss the uncertainties in the quark mass, coming from different
choices of the expansion parameter in sec. \ref{sec:bk}.
\par  We can give an alternative
definition of the renormalized  mass, which is regularization-independent.
One can for example define the renormalization constant
of the  mass $Z_m$ as the inverse of the renormalization constant of the
scalar density $S=\bar \psi \psi$:
 \be Z_m=Z_S^{-1},  \label{eq:zms}\ee
where $Z_S$ is defined by the renormalization condition:
\be Z_S(\mu a)  \Gamma_S(\mu a) / \Gamma^0_S= 1 \label{eq:rcrc} \ee
imposed to the amputated Green function
on off-shell quark states with $p^2=\mu^2$.
Since $Z_S$ in eq. (\ref{eq:rcrc}) is gauge-dependent, we will
denote it as $Z_S^{gauge=lan, fey,\dots}$
in the following. The renormalization scheme
introduced in eq. (\ref{eq:rcrc}),
which we call in the following $RI$ for regularization-independent,
 is particularly
suitable for the implementation of the relevant Ward identities of the
theory, see sec. \ref{sec:wids}, and for the non-perturbative renormalization
of lattice operators \cite{npr}. In perturbation
theory, in the Feynman or Landau gauge, one finds:
\bea m^{fey,lan}(\mu)
&=& \Bigl(\frac{\as(\mu)}{\as(\pi/a)}\Bigr)^{\gamma^{(0)}/2 \beta_0}
\Bigl[ 1+ \frac{\as(\mu)-\as(\pi/a)}{4 \pi}
\Bigl(  \frac{\gamma^{(1)}}{2\beta_0}-
\frac{\gamma^{(0)}\beta_1}{2\beta_0^2})
\nn \\
&+&\frac{\as(\pi/a)}{4 \pi} K_m- \frac{\as(\mu)}{4 \pi}
 C^{fey,lan}_m \Bigr] m(a)  \nn \\
&\simeq& \Bigl[1 - \frac{\as(\mu)}{4 \pi}
 C^{fey,lan}_m \Bigr] \msms \nn \\
 C^{fey}_m&=&-\cf  5 \nn \\
 C^{lan}_m&=&-\cf  4 \label{eq:msren1} \eea
As explained in ref. \cite{ciuc}, $m^{fey,lan}(\mu)$ are
regularization-independent.
In fact they are obtained by imposing the same renormalization
conditions in all the  regularizations. The price  is that
this definition  can be gauge-dependent
 and one has to specify the external states at which the
renormalization conditions have been imposed.
\subsection{ Tadpole resummation}
\label{sec:ms2}
In the Wilson or SW--Clover lattice formulation of the quark action,
$m(a)$ is expressed in terms of the hopping parameter as:
\be m(a) = \ln\Bigl(1+\frac{1}{2}( \frac{1}{K} - \frac{1}{K_c^0}) \Bigr)
\simeq \frac{1}{2}\Bigl( \frac{1}{K} - \frac{1}{K_c^0} \Bigr), \ee
where the critical value of $K$, corresponding to $m(a)=0$, is $K_c^0=1/8$.
At first order in perturbation theory, $K_c$ is given by:
\be \frac{1}{K_c}=8-2 {\Sigma_0^L}=8 u_0 . \ee
It has been argued that lattice perturbation  theory is dominated by
tadpole diagrams, which have no correspondence in the continuum,
and that the hopping parameter should be replaced by an ``effective"
one \cite{lm1,lm2}:
\be \bar K = K u_0 \sim \frac{K}{8 K_c}, \ee
where $K_c$ is defined non-perturbatively as the value of $K$ at which
the pseudoscalar meson mass vanishes.
The same tadpole diagrams contributing to $\Sigma_0^L$ also enter  the
calculation of $\Sigma_1$. At first order in perturbation theory, we
can write:
\bea 1 - \Sigma_1^L= 1-\frac{\Sigma_0^L}{4}-(\Sigma_1^L-\frac{\Sigma_0^L}{4})
\nn \\ \simeq (1-\frac{\Sigma_0^L}{4})(1-\bar \Sigma_1^L)=
\frac{1}{8 K_c}(1-\bar \Sigma_1^L). \label{eq:tp}\eea
Using eq. (\ref{eq:tp}),
the lattice inverse quark propagator  becomes:
\bea \is&=&  \frac{\Bigl[1-\bar \Sigma_1^L(pa,\bar m(a) a)\Bigr]}{8 K_c}
\nn \\ &\times& \Bigl[
i p\!\!\!/ + \bar m(a)\Bigl(1+\bar \Sigma_1^L(pa, \bar m(a)a)
- \Sigma_2^L(pa,\bar m(a)a)\Bigr)\Bigr] \label{eq:qp2} \eea
where $\bar m(a) = 4K_c/K-4 \simeq \ln(4 K_c/K-3)$.
The relation between the lattice quark mass $\bar m(a)$
and $\msms$ is then given by the same expression as in eq. (\ref{eq:msren}),
with $\bar m(a)$ instead of $m(a)$ and $\bar \sigma_1=\sigma_1-\Sigma_0^L/4$
 instead
of $\sigma_1$ (see table \ref{tab:sigs}).
\subsection{{\bf Renormalization conditions and Ward identities}}
\label{sec:wids}
In this section we discuss the definition of the quark mass via the
Ward  identities of the regularized theory
and its relation with the quark
mass defined in the previous section\footnote{ To
 avoid problems connected with anomalous terms, we
will  only consider Ward indentities for non-singlet currents.}.
\par In the continuum, the renormalization of the quark mass
and of the scalar density are related by the vector
current Ward identity:
\be \langle  \alpha \vert \partial^\mu V^a_\mu
\vert \beta \rangle =\langle  \alpha \vert \partial ^\mu \bar \psi \gamma_\mu
\frac{\lambda^a}{2}\psi \vert \beta \rangle= \langle  \alpha \vert
\bar   \psi \left[ M,\frac{\lambda^a}{2} \right]\psi  \vert \beta \rangle
\label{eq:cvc} \ee
where the $\psi$'s are bare fields, $M$ is the bare mass matrix
and $ \vert \alpha,\beta \rangle$ are arbitrary on-shell physical states.
Equation (\ref{eq:cvc}) guarantees that the
product $ m \bar \psi \psi$ is unrenormalized by strong interactions:
\be m_{R} \bar \psi_{R} \psi_{R} = m Z_m Z_s \bar \psi \psi
, \label{eq:eqm} \ee
i.e. that $Z_m=Z_S^{-1}$ as in eq. (\ref{eq:zms}), which remains
valid on the lattice, where it is also possible to define a conserved vector
current in the  limit of degenerate quark masses. It ensures that the
mass and scalar density renormalization constants
are the inverse of one another to all orders in perturbation
theory and beyond. Still we have the freedom to decide which definition of
the renormalized mass we wish to use: renormalized
 on the mass-shell, in the $\overline{MS}$
(gauge-invariant)  or in the $RI$ (gauge-dependent) schemes,
 as explained in sec. \ref{sec:ms}. Of course we could decide
to renormalize the mass in the minimal-subtraction scheme and the
scalar density in $RI$. Such an exotic choice would only obscure the
understanding of the Ward identities and will not be considered here.
Differences between different definitions of the quark mass
are related by terms computable in perturbation theory. This is true for
heavy quarks with the renormalization on the mass shell and for
heavy or light quarks with the $\overline{MS}$ or  $RI$ renormalizations,
provided that $\mu$ and $\pi/a$ are much larger than $\Lambda_{QCD}$.
\par The quark mass can also be defined through the Ward identity of the
axial current:
\be  \langle  \alpha \vert
\partial^\mu A^a_\mu  \vert \beta \rangle =  \langle  \alpha \vert
\partial ^\mu \bar \psi \gamma_\mu \gamma_5
\frac{\lambda^a}{2} \psi  \vert \beta \rangle =
  \langle  \alpha \vert
 \bar \psi \left\{ M,\frac{\lambda^a}{2} \right\} \gamma_5 \psi
 \vert \beta \rangle. \label{eq:pcac} \ee
In a regularization that preserves chirality\footnote{ This
regularization strictly speaking does
not exist.},
as for example na\"\i ve $\overline{MS}$, the scalar and
pseudoscalar ($P^a=\bar \psi\lambda^a/2 \gamma_5 \psi$) densities have the
same renormalization constants, $Z_S=Z_P$, so that the quark mass defined
from eq. (\ref{eq:cvc}) or (\ref{eq:pcac}) coincide. This is not true
in continuum regularizations, which break chirality, like 't Hooft--Veltman
or dimensional reduction in the continuum,
 or in the Wilson (SW--Clover) formulation of
QCD on the lattice,
where an explicit term that violates chirality is present in
the action. We now explain how the quark mass can be defined through the
axial Ward identity in the presence of an explicit breaking of chiral
invariance. The argument is made with the lattice regularization,
but is general.
\par At zero order in $\as$
the axial current Ward identity on the lattice can be written
as \cite{boc,ks}:
\be   \langle  \alpha \vert\partial^\mu A^a_\mu  \vert \beta \rangle=
2 m    \langle  \alpha \vert  P^a  \vert \beta \rangle
+  \langle  \alpha \vert \xi^a_A  \vert \beta \rangle \label{eq:pcacl} \ee
in terms of bare fields and masses. We have simplified
the Ward identity by taking degenerate quark masses;
 $\xi^a_A$ is the chiral rotation of
the Wilson term, which is of $O(a)$ only in the free-field case
($\sim a \bar \psi D^2 \psi$).
Because of interaction, $\xi^a_A$ develops terms of $O(g_L^2/a)$ and
$g_L^2$ \cite{boc}, which renormalize the mass term, the pseudoscalar density
and the axial current. Equation (\ref{eq:pcacl}) becomes then:
\be Z_A   \langle  \alpha \vert\partial^\mu A^a_\mu  \vert \beta \rangle=
2 \Bigl( m-\bar m(m) \Bigr)  \langle  \alpha \vert P^a \vert \beta \rangle
+   \langle  \alpha \vert \bar \xi^a_A  \vert \beta \rangle,
\label{eq:pcacll} \ee
where $\bar \xi^a_A$ is of $O(a)$, including strong-interaction effects.
Close to the chiral limit, and neglecting terms of $O(a)$,
eq. (\ref{eq:pcacll}) becomes:
\bea Z_A   \langle  \alpha \vert \partial^\mu A^a_\mu \vert \beta \rangle& =&
2 \Bigl( m- m_c \Bigr)\Bigl(1-\frac{\partial \bar m}{\partial m_c}
\Bigr)   \langle  \alpha \vert P^a \vert \beta \rangle \nn \\ &=&
2 \Bigl( m- m_c \Bigr)
\frac{Z_P}{Z_S}   \langle  \alpha \vert P^a \vert \beta \rangle,
\label{eq:hic}  \eea
where the chiral value of the mass is defined by the equation
$m_c=\bar m(m_c)$  and corresponds to $K=K_c$ in the notation used
in sec. \ref{sec:ms},  $m-m_c \simeq 1/2(1/K-1/K_c)$.
The factor $Z_P/Z_S$ appears since $S$ and $P^\prime=Z_P/Z_S P$
(and not $S$ and $P$ \cite{boc})
belong to the same chiral multiplet, because of the symmetry breaking
present on the lattice. The same would be true with the 't Hooft--Veltman
regularization.
\par Even though  $Z_A$ and $Z_P/Z_S$ can be both computed
in perturbation theory, an alternative, fully
non-perturbative,  definition of the quark mass can be given.
Using the Ward identities relative to the axial vector and
scalar--pseudoscalar densities we can determine both $Z_A$ and
$Z_P/Z_S$ \cite{npr,vla,mm,pac} in a non-perturbative way.
Then we can measure the ratio $2 \rho$ of the matrix elements of
$\partial^\mu A^a_\mu$ and $P^a$
on a physical state; typically one uses:
\be 2\rho = \frac{  \langle 0 \vert \partial^0 A^a_0 \vert \pi(\vec p=0)
 \rangle}
{  \langle 0 \vert P^a \vert \pi(\vec p=0) \rangle}\label{rho} \ee
and then, by saturating the Ward identity in eq. (\ref{eq:hic}) we
can find the lattice bare quark mass. This avoids ambiguities
on the mass definition of the kind $m(a)=4(K_c/K-1) \sim \ln(4 K_c/K-3)$, see
sec. \ref{sec:ms2} \footnote{ There remains another ambiguity of $O(a)$ in the
definition of the time derivative in eq. (\ref{rho}). This ambiguity can
be made of $O(\as a)$ with improved actions.}.
In practice we define $m=m(a)$ as:
\be m(a)= Z_A \times \frac{Z_S}{Z_P} \times \rho, \nn \ee
\par To find the continuum, renormalized quark mass, we can proceed
at this point in two different ways. We can work in perturbation theory,
and we essentially recover the results found in eq. (\ref{eq:msren}) of
sec. \ref{sec:ms2}. Alternatively we can renormalize non-perturbatively
 the pseudoscalar
density by imposing, on quark states of momentum
$p^2=\mu^2$ and in a fixed (Landau) gauge, the
renormalization conditions \cite{npr}:
\be  Z^{lan}_{S}(\mu a) \Gamma_S(\mu a) / \Gamma^0_S=
Z^{lan}_{S}(\mu a ) \Gamma_{P^\prime}(\mu a) / \Gamma^0_{P}=
Z^{lan}_{P}(\mu a) \Gamma_{P}(\mu a) / \Gamma^0_P= 1. \nn \ee
We have used the fact that the renormalization conditions imposed
on $S$ and $P$ may be chosen in such a way as
to satisfy the continuum Ward identities, i.e.
$Z_P/Z_S=Z_P^{lan}/Z_S^{lan}$.
The continuum mass is then given by:
\be m^{lan}(\mu)= Z^{lan}_m(\mu a) m(a)=
Z_A \times \bigl(Z_P^{lan}(\mu a)\Bigr)^{-1}
 \times \rho , \label{eq:mlan} \ee
We can then use eq. (\ref{eq:msren1})   to obtain $\msms$
from $m^{lan}(\mu)$.
The advantage of this procedure is that perturbation theory is
used only to relate  quark masses in the continuum, thus avoiding
the large perturbative corrections present on the lattice \cite{npr}.
The possible disadvantage is that
$\mu$ has to satisfy the condition $\Lambda_{QCD} \ll \mu  \ll 1/a $
to avoid large higher-order corrections and discretization
errors\footnote{ One could imagine determining
$Z_{S,P}^{lan}$  non-perturbatively
on the lattice, at very large values of $\beta$, thus avoiding large
higher-order effects. This however would not solve the problem because, in
order
to get a physical quantity from a matrix element computed in current
simulations, one has in any case to evolve the operators down to a scale
$\mu$ smaller than the inverse lattice spacing used in the numerical
 calculation of the matrix element. Thus large higher-order effects would
be present anyway.}.
It turns out that, using the SW--Clover action at $\beta=6.0$, it is possible
to choose a value of $\mu \sim 1/a \sim 2$ GeV, at which
 good agreement can be found between the determination of $Z^{lan}_P/Z^{lan}_S$
on quark states \cite{npr}
 and $Z_P/Z_S$ as computed using the Ward identities \cite{vla},
with a relatively small discretization error.
The  systematic error due to discretization can be estimated
to be of the order of $\sim 10$--$15\%$.
Since at present $Z^{lan}_P$ has been computed only at $\beta=6.0$
and only with the SW--Clover action, the results of the non-perturbative
method will be reported in the
next section, but they  will not be included  in the final evaluation of
$m_s$.
\begin{figure}[t]
  \begin{center}
     \setlength{\unitlength}{1truecm}
     \begin{picture}(6.0,6.0)
        \put(-5.0,-6.2){\special{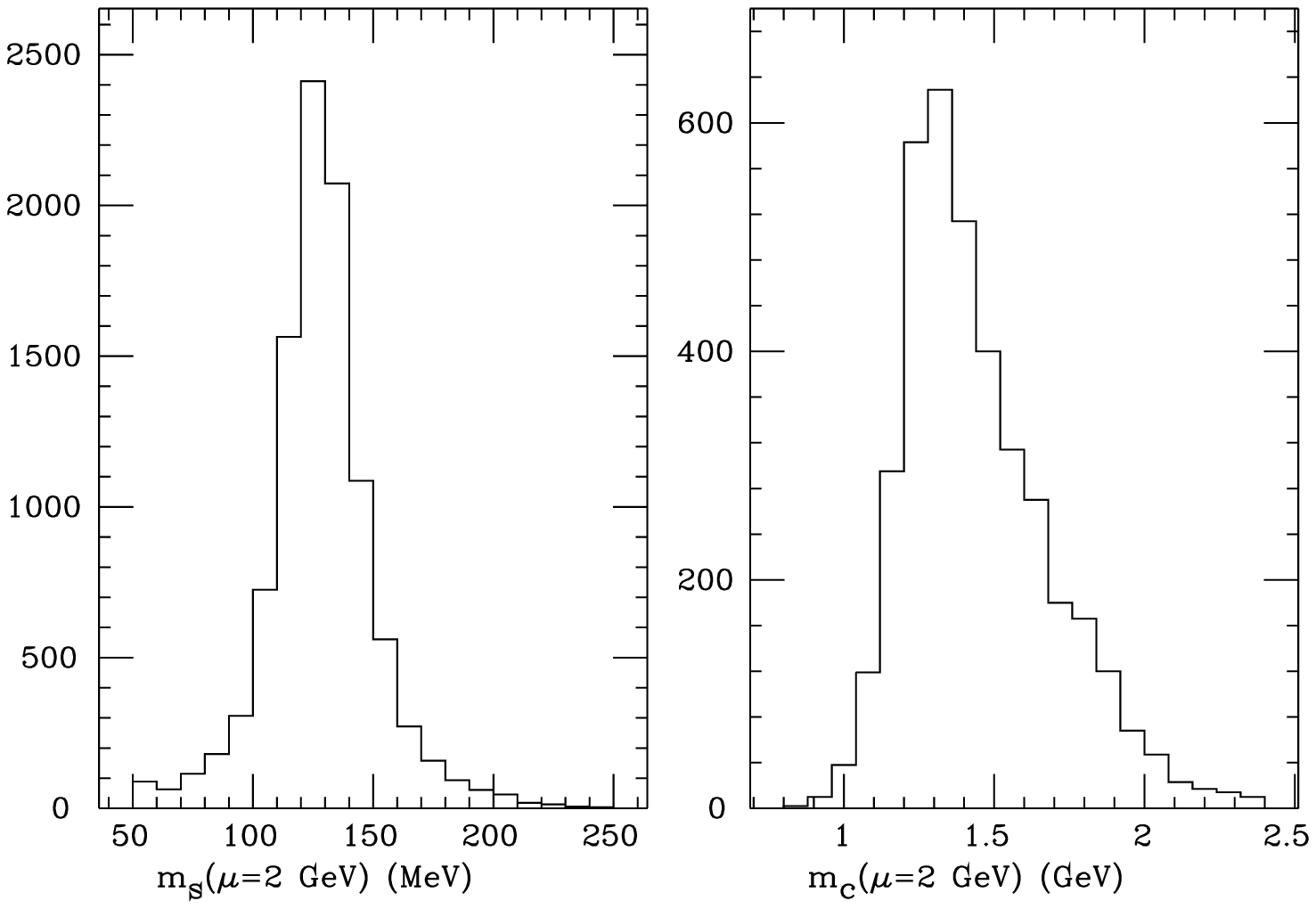}}
     \end{picture}
  \end{center}
  \caption[]{\it{Histograms of the strange and charm quark mass distributions
in arbitrary units,
computed
in the $\overline{MS}$ scheme at the scale $\mu=2$ GeV. They have been obtained
by  considering all possible uncertainties discussed in the text.
  In the case of the strange quark
all the results  of table \ref{tab:ms} have been included. In the
charm case, in order to reduce discretization errors,
we have only considered the results obtained at $\beta \ge 6.2$.}}
  \protect\label{fig:isto}
\end{figure}
\begin{figure}[t]
  \begin{center}
     \setlength{\unitlength}{1truecm}
     \begin{picture}(6.0,6.0)
        \put(-5.0,-6.2){\special{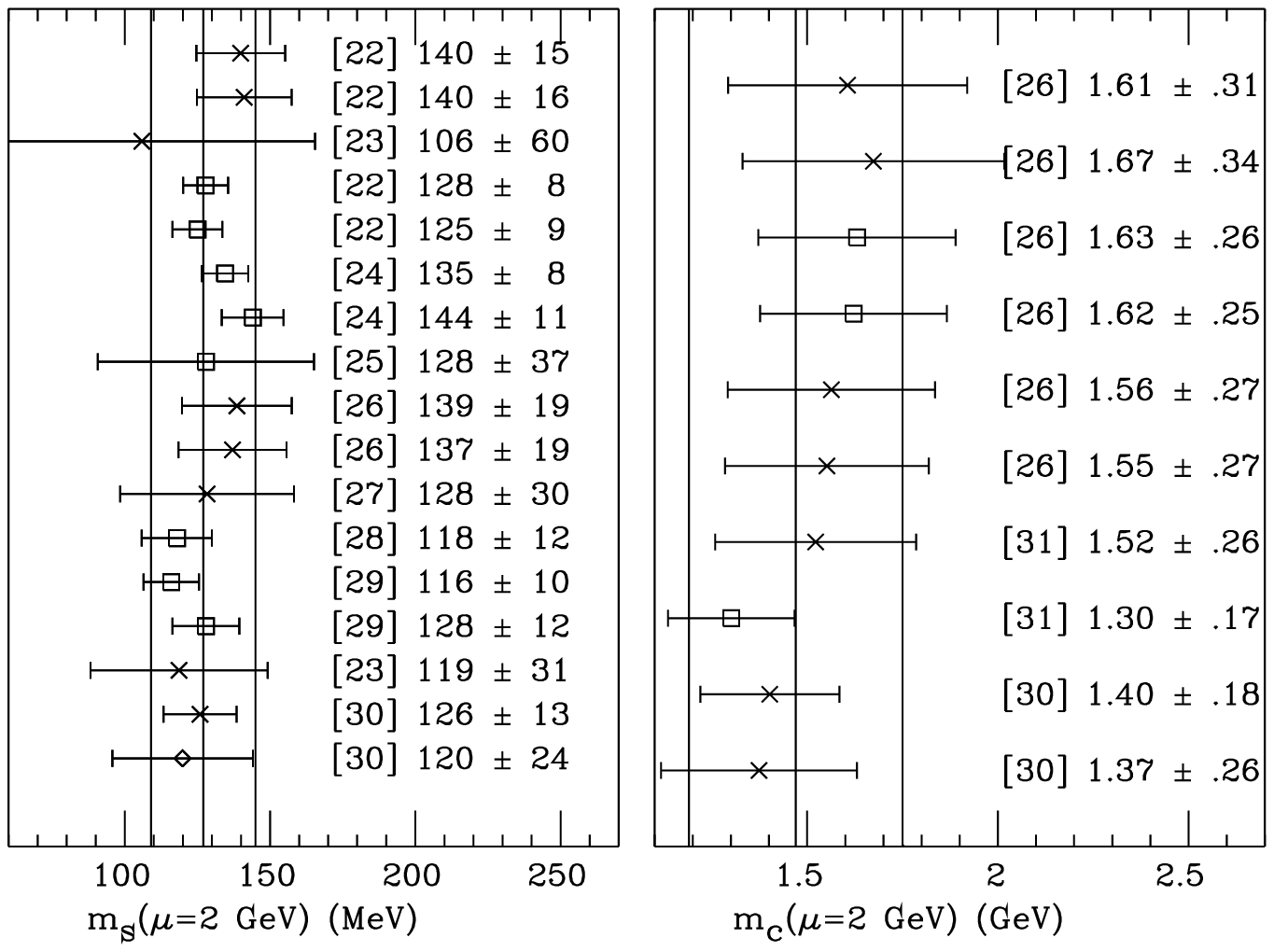}}
     \end{picture}
  \end{center}
  \caption[]{\it{Strange and charm $\overline{MS}$ masses and errors
at $\mu=2$ GeV from the results of  table \ref{tab:ms}.}}
  \protect\label{fig:all_lat}
\end{figure}
\section{Numerical estimates of quark masses from the lattice}
\label{sec:bk}
\subsection{{\bf Uncertainties of the lattice calculations}}
We are now ready to give the results for the continuum quark masses.
These results have been obtained by combining the values of the bare lattice
masses given in table \ref{tab:ms} with the conversion factor in
eq. (\ref{eq:msren}), computed
with all its possible allowed variations, value of the scale,
$\as$, etc.
The theoretical predictions are subject to the uncertainties  listed
below:
\begin{itemize}
\item {\bf Statistical error on $m(a) a$}: this error is usually
in the range $5$--$20 \%$ depending on the value of $\beta=6/g_L^2(a)$
 and on the statistical accuracy of the simulation.
\item {\bf Calibration of the lattice spacing}: the uncertainty on the
value of $a^{-1}$ in physical units enters in two ways:
\par
i) when we convert
the dimensionless quark mass $m(a) a$, which is directly accessible in
 numerical simulations, to $m(a)$;
\par
ii) in the scale used to evaluate
$\as(\pi/a)$ or $\as(1/a)$, etc.
\par
Only the first effect is important,
since the second one gives  a mild, logarithmic dependence in
$a$. The value of $a$ enters not only
in the conversion of $m(a) a$ to $m(a)$, but also in
the determination of $K_s$ ($K_{ch}$) from the mass of a strange
(charmed) hadron.
 There are several methods to evaluate the lattice
spacing. The most popular ones are
from the string tension, from the mass of the $\rho$, and from $f_\pi$.
They  usually differ  from one another  by $\sim 10\%$.
\item {\bf Choice of the expansion parameter}: we can use $\as$,
$\as^{V_1}$ or $\as^{V_2}$, or any other coupling constant that
one believes makes the perturbative series converge rapidly.
The differences are of $O(\as^2)$, but may be important with the
lattice regularization \cite{lm1,lm2}.
We have evaluated the conversion factor with all the three
possibilities listed above. In particular, with the $\overline{MS}$
choice of $\as$, we have evaluated $\as$ either by using
eq. (\ref{eq:asms})
or by using the ``unquenched" LEP results \cite{altas,german}, i.e. we have
computed $\as$ from eq. (\ref{eq:srcc}) with $\Lambda_{QCD}^{n_f=4}
= (340 \pm 120)$ MeV.
\item {\bf Effects of $O(\as^2)$}: Values
of the mass that differ by terms of  order $\as^2$ are obtained by using
the two equivalent versions of eq. (\ref{eq:cmw1}),
but with different scales
as arguments of $\as$. We have allowed the scale in $\as$ to vary
between $\pi/a$ and $1/a$, which seems to us sufficient to cover a large
range of possibilities.
\item {\bf Tadpole-improved definition of the mass}:
We have evaluated the continuum mass from $m(a)$ defined as $1/2(1/K-
1/K_c)$ and from its tadpole improved expression $\ln(4 K_c/K-3)$ with
the conversion factor changed accordingly.
\item {\bf Discretization errors}: In the final estimate of the
mass we combine  the results obtained with the Wilson
and SW--Clover ``improved" action, which suffer from different $O(a)$ effects.
The comparison is useful because in the SW--Clover case, discretization
errors are of $O(\as a)$ and have been found in some cases as small
as $5\%$, to be compared to $30 \%$ in the Wilson case.
\item {\bf Quenching}: It is not possible to estimate and correct the
error due to the quenched approximation. We have only applied  the
 following procedure. From $m(a)$ one can get $m^{\overline{MS}}(\pi/a)$,
which is subsequently  evolved to $\mu=2$ GeV. It is clear that
$m^{\overline{MS}}(\pi/a)$ scales in $a$
(and for consistency in  $\mu$)  with a quenched anomalous dimension
and $\beta$-function.  However, by imposing $\as(\pi/a)^{quenched}=
\as(\pi/a)^{unquenched}$,  $m^{\overline{MS}}(\pi/a)$ has been evolved also
with the unquenched formulae and the  difference has been taken into account
in the final error. This procedure does not pretend to correct for the
systematic error entailed by the quenched approximation, but it is
consistent with the observation that most of the quenched and unquenched
results
look very similar  after a suitable rescaling of the lattice coupling
constant.
\end{itemize}
\subsection{{\bf Results for the strange and charmed quark masses}}
\par By changing the value of the lattice spacing, the expansion parameter
$\as$, the scale at which $\as$ is evaluated, etc., and taking into
account the statistical errors of the lattice simulations,
we obtain a pseudo-Gaussian distribution of the value
of the mass from which it is possible to evaluate the theoretical error,
following the method explained in ref. \cite{upgraded}, see fig. 1.
The continuum determination, at the scale $\mu=2$ GeV,
of  masses and relative errors from different lattice simulations are
reported in figs. \ref{fig:isto} and \ref{fig:all_lat}.
One notices a nice agreement
between different simulations performed at different values of the lattice
spacing and with different lattice quark actions. No systematic trend
in $\mu$ or dependence on the lattice action appears for the strange quark.
In the case of the charm quark, a slight reduction of the value of the
mass seems to appear at larger values of $\beta$. This could be due
to terms of $O(m_{ch} a)$. In order to reduce the systematic error due
to these corrections, in the final estimate of the value of the mass,
we have taken only the determinations of $m_{ch}$ at $\beta \ge 6.2$.
 Combining all the results together, our
best estimates, which is also reported in the
abstract, are given by:
\bea m_s^{\overline{MS}}(\mu=2\,\, \rm{ GeV})
&=& (128 \pm 18)\,\, \rm{MeV} \nn \\
 m_{ch}^ {\overline{MS}}(\mu=2 \,\,\rm{ GeV})&=& (1.48 \pm 0.28)\,\, \rm{GeV}
. \label{eq:mmu}\eea
\par For completeness we also try to estimate $m_s$, with the SW--Clover
action,
using the non-perturbative method envisaged in sec. \ref{sec:wids}.
{}From a recent study at $\beta=6.0$ they found \cite{all3}:
\be 2\rho(K=K_s)= 0.071(5),  \,\,\,\,\,\, \beta=6.0 . \ee
{}From the  non-perturbative lattice renormalization of $P^a$, at
$\beta=6.0$ and $(pa)^2 \sim 1$, they estimated \cite{npr}:
\be Z_P^{lan}(\mu a \sim 1)= 0.49(1)\ee
from which, using eq. (\ref{eq:mlan}), one finds:
\be m^{lan}_s(\mu \sim 2 {\rm GeV})= (150 \pm 18 )\,\, \rm{MeV},
\label{npr1}\ee
or, from  relation (\ref{eq:msren1}):
\be m_s^{\overline{MS}}(\mu=2{\rm GeV})=(141 \pm 17 )\,\, {\rm MeV}
, \label{npr2} \ee
in reasonable  agreement with the results of eq. (\ref{eq:mmu}).
In eqs. (\ref{npr1}) and (\ref{npr2}) only
the statistical  errors are reported.
A further uncertainty of $\sim 10-15 \%$, due to discretization errors, must be
taken into account. A more accurate analysis can be found in ref. \cite{npr}.
\section{Conclusion}
We have shown that lattice simulations of QCD can give accurate and
reliable  determinations of quark masses, using well-defined procedures.
At present, it is possible to evaluate most of the uncertainties with the
only exception of
the systematic error introduced by the quenched approximation.
The values of the strange and charm quark masses found in this work can
readily be used for phenomenological applications.
\section*{{\bf Acknowledgements}}
We warmly thank A.L. Kataev, C.T. Sachrajda, L. Silvestrini,
M. Testa, and A. Vladikas for discussions.
We thanks all the members of the APE collaboration for the use of some
unpublished results.
We acknowledge the partial support of the MURST, Italy, and  the INFN.

\end{document}